\documentstyle[eqsecnum, prl,aps,epsfig,amscd]{revtex}

\begin{document}
\draft
\twocolumn

\def\up{|\uparrow\,\rangle}
\def\dn{|\downarrow\,\rangle}
\def\upd{\langle\, \uparrow |}
\def\dnd{\langle\, \downarrow |}
\def\upup{|\uparrow\,\uparrow \rangle}
\def\dndn{|\downarrow\,\downarrow \rangle}
\def\updn{|\uparrow\,\downarrow \rangle}
\def\dnup{|\downarrow\,\uparrow \rangle}
\def\upupd{\langle \uparrow\,\uparrow |}
\def\dndnd{\langle \downarrow\,\downarrow |}
\def\updnd{\langle \uparrow\,\downarrow |}
\def\dnupd{\langle \downarrow\,\uparrow |}
\def\beqn{\begin{equation}}
\def\eeqn{\end{equation}}
\def\beqnar{\begin{eqnarray}}
\def\eeqnar{\end{eqnarray}}
\newcommand{\tfrac}[2]{{\textstyle\frac{#1}{#2}}}
\newcommand{\mb}[1]{\mbox{\boldmath{$#1$}}}
\def\ba{\begin{array}}
\def\ea{\end{array}}
\newcommand{\etal}{{\em et al. }}
\newcommand{\eqn}[1]{(\ref{#1})}
\newcommand{\arowspace}[1]{\renewcommand{\arraystretch}{#1}}
\newcommand{\acolspace}[1]{\renewcommand{\arraycolsep}{#1}}

\preprint{HEP/123-qed}
\title{Quantum simulation of a three-body interaction Hamiltonian \\on an NMR quantum computer}

\author{
C. H. Tseng{$^1$}, 
S. Somaroo{$^1$}, 
Y. Sharf{$^1$}, 
E. Knill{$^2$}
R. Laflamme{$^2$}, 
T. F. Havel{$^3$}, 
and D. G. Cory{$^1$}\footnote{
Correspondence and requests for materials should 
be addressed to D. G. C. (email:{\it dcory@mit.edu}).}
}

\address{
{$^1$}Department of Nuclear Engineering, Massachusetts Institute of Technology, Cambridge, MA 02139\\
{$^2$}Theoretical Physics Division, Los Alamos National Laboratory, Los Alamos, NM 87455\\
{$^3$}BCMP Harvard Medical School, 240 Longwood Avenue, Boston MA 02115
}

\date{\today}
\maketitle
\begin{abstract}

Extensions of average Hamiltonian theory to quantum computation permit the design of 
arbitrary Hamiltonians, allowing rotations throughout a large Hilbert space.   
In this way, the kinematics and dynamics of any quantum system may be simulated by 
a quantum computer.  A basis mapping between the systems 
dictates the average Hamiltonian in the quantum computer needed to implement the desired Hamiltonian in the 
simulated system.  The flexibility of the procedure is illustrated with NMR on ${\rm ^{13} C}$ labelled Alanine by creating the 
non-physical Hamiltonian ${\sigma_{z}\sigma_{z}\sigma_{z}}$ corresponding to a three body interaction.  
\end{abstract}


\narrowtext
\section{Introduction}
In the early 1980's, researchers such as Benioff, Bennett, Deutsch, Feynman and Landauer \cite{Benioff,Bennett,Deutsch,Feynman,Landauer} studied the possibility of performing computations using the principles of quantum mechanics, and conjectured that a machine based on these principles might be able to solve certain types of problems more efficiently than can be done on a conventional von Neumann computer. Later, Lloyd proposed that such a quantum computer might be built from an array of coupled two-state quantum systems, each of which can store one quantum bit, or qubit, of information \cite{Lloyd1,Lloyd2}. Shortly thereafter, Shor proved that a quantum computer would be capable of factorizing integers in polynomial time, thereby showing that the exponential number of degrees of freedom accessible to a quantum computer does indeed enable it to solve some problems more efficiently than is believed possible on a conventional machine \cite{Shor}. 

An essential feature of a universal quantum computer is the ability to transform efficiently any initial state vector to any other state vector within a large Hilbert space. Such operations can be thought of as using a real-number (continuous) rather than a Turing-machine (digital) model of computation \cite{Traub}.  Algorithms can be tailored to take advantage of this as well as the parallelism from quantum superposition. 
Since any quantum system governed by an arbitrary Hamiltonian may be described by the paths taken by a set of basis vectors, a quantum information processor should be able to simulate the evolution of any smaller quantum system up to a specified time point.  

Accordingly, one of the first proposed applications of quantum computers was quantum simulation: using a quantum mechanical computer to simulate another quantum mechanical system.  Feynman's original proposal in 1982 \cite{Feynman} that there might be a universal and efficient quantum simulator of physical systems was recently validated, in general terms, by Lloyd \cite{Lloyd3}.  Algorithms for specific classes of quantum systems have also been proposed \cite{Weisner,Zalka,Abrams,Lidar}.  To apply a quantum algorithm like factorization to a problem beyond the reach of conventional computers, a quantum computer will have to perform millions of operations coherently and substantially without error. Although such control can in principle be achieved through quantum error correction \cite{Knill,Knill2,Cory,Steane,Braunstein,Calderbank,Zurek}, this involves a very high overhead.  In contrast, useful quantum simulations will perhaps require only hundreds of operations.  

While it is possible to simulate a quantum system on a classical computer, it becomes increasingly difficult as the size of the system increases to store the quantum state, much less to compute the entire wavefunction evolution.  For example, a quantum system of 50 spin one-half particles occupies a Hilbert space of dimension $2^{50} \sim 10^{15}$.  This requires $\sim 10^{15}$ complex numbers to specify the state completely.  While intractable using classical computers, a quantum mechanical device would require only 50 qubits to store the state of the system.  Evolving this state vector is also difficult.  For the case of local interactions (all systems obeying special and general relativity), Lloyd has suggested an efficient construction of the evolution operator with small time steps of evolution under local interactions \cite{Lloyd3}.

This article describes a general scheme for implementing quantum simulations, and illustrates the flexibility of the method with the synthesis of a Hamiltonian not normally found in Nature. The challenge that remains is to find a sequence of propagators or ``gates'' that can {\it efficiently} produce the desired behavior in practical problems. 

\section{Quantum Simulation}
Many of the concepts of quantum simulation are 
implicit in the average Hamiltonian theory (AHT) developed by Waugh and colleagues to design NMR pulse sequences which implement a specific desired effective NMR
 Hamiltonian \cite{Waugh}.  While AHT as applied to NMR spectroscopy is often directed at obtaining a selectively scaled version of the internal Hamiltonian ${\cal H}_{int}$, the formalism makes clear that other Hamiltonians in the operator space may be constructed.  The significant contribution to quantum information processing is to articulate the range of propagators $U$ that may be simulated given ${\cal H}_{int}$ and allowed external interactions. Through the tenets of quantum information processing it is clear that, provided there are coupling pathways (interactions) connecting any two identifiable qubits then any Hamiltonian that spans this space may be constructed \cite{Lloyd3,Barenco,Divincenzo2}.  AHT provides a means to quantify the precision of a specific simulation and allows a systematic improvement of the precision of the implementation.  

A general scheme for quantum simulation utilizing some of the results of AHT is based on establishing a correspondence between a simulated (model) system, S, and a physical (experimental) system, P \cite{Somaroo}.  This is summarized by the following diagram:

\begin{equation}
\ba{ccc}
|s\rangle  			&\stackrel{\phi}{\longrightarrow}  	&|p\rangle 		\nonumber\\
&&\vspace{0.2pt}\nonumber\\
{U} \downarrow	\; 		& 					&\; \downarrow {V_T} 	\nonumber\\
&&\vspace{0.2pt}\nonumber\\
|s(T)\rangle 			&\stackrel{\phi^{-1}}{\longleftarrow}   &|{p_{T}}\rangle \nonumber
\ea
\label{scheme}
\end{equation}
The goal is effect the evolution of the simulated system $|s\rangle \stackrel{U}{\longrightarrow} |s(T)\rangle$ using the physical 
system $P$, where the propagator $U=e^{-i{\cal H}_sT/\hbar}$, and ${\cal H}_s$ is the desired Hamiltonian governing the simulated system.  To do this, $S$ is related to $P$ by an invertible map $\phi$ 
which determines a correspondence between all the operators and states of $S$ 
and $P$. In particular, the propagator $U$ maps to $V_T = \phi^{-1} U \phi$.
The challenge is to implement $V_T$ using propagators $V_i$ arising from the available external interactions with intervening periods of natural evolution 
$e^{-i {\cal H}_p^{0} t/\hbar}$ in $P$ so that 
$V_T = {\Pi}_{i} e^{-i {\cal H}_p^{0} t_i(T)}V_i$. 
If a sufficient class of simple operations (logic gates) are implementable in 
the physical system, any operator (in particular 
$V_T$) can be composed of natural evolutions in $P$ and external interactions \cite{Lloyd3,Divincenzo2,Shoretal,Divincenzo,DivincShor}.
For unitary maps $\phi$, we may write $V_T = e^{-i\overline{\cal H}_p T/\hbar}$
where $\overline{\cal H}_p \equiv \phi^{\dag}{\cal H}_s\phi$ can be identified 
with the average Hamiltonian introduced by Waugh \cite{Waugh,Waugh2}. 
After 
$|p\rangle \stackrel{V_T}{\longrightarrow} |p_{T}\rangle $, the final map 
$\phi^{\dag}$ takes $|p_{T}\rangle \rightarrow |s(T)\rangle$ thereby effecting 
the simulation $|s\rangle \rightarrow |s(T)\rangle$.
Note that ${\cal H}_s(T)$ can be a time dependent Hamiltonian and that $T$ is 
treated as a parameter when mapped to $P$. This means that the physical times
$t_i(T)$ are parameterized by the simulated time $T$. 

The desired simulated Hamiltonian may be specified in various ways.  On the one hand if the simulated Hamiltonian is specified by eigenenergies, translation to a representation in terms of the Pauli matrices $\sigma_z^i$ may provide further physical insight and facilitate implementation via geometric algebra techniques \cite{GApaper} using a weakly coupled spin system.  Although no general compiler from a Pauli matrix expression is known, the universality of a quantum computer implies that an implementation exists.  Examples of such Hamiltonians include the Balmer series of the hydrogen atom and the harmonic oscillator.  On the other hand, if the simulated Hamiltonian is specified in terms of Pauli matrices, translation to an eigenstructure representation permits experimental verification of its spectral structure.  Examples include the Ising model ${\cal H}_s = \Sigma_{i<j}  \alpha_{ij} \sigma_z^i \sigma_z^j$ and the three-body interaction ${\cal H}_s = {\hbar \over 2}\pi J_{123}\sigma_{z}^1\sigma_{z}^2\sigma_{z}^3$ discussed in this paper. A general method for systems with known eigenstructure is given in the Appendix.  For cases where the eigenstructure is not known or the Hamiltonian is not expressed in terms of the $\sigma_z^i$, simulation is still possible.  For example, in the treatment of the driven anharmonic oscillator in Ref.\cite{Somaroo} the eigenstructure is not assumed.

\section{Implementation of Effective  ${\bf \sigma_{z}\sigma_{z}\sigma_{z}}$ Hamiltonian }

Liquid state NMR quantum computers \cite{Cory2,Gershenfeld,Steane2} are well suited for quantum simulations 
because they have long relaxation times ($T_1$ and $T_2$) as well as the 
flexibility of using a variety of molecular samples. In particular, the 
`scalar' coupling between the nuclear spins, denoted $J$, may be reduced at will by means of radiofrequency pulses. Typically spin 1/2 nuclei are used.  Thus, the kinematics of any $2^N$ level quantum system could be simulated using a given 
$N$-spin molecule.  We will use a 3-spin system, and illustrate the flexibility of the scheme by implementing a non-physical three body interaction (see also \cite{Bruschweiler}).  For a weakly coupled system ${\cal H}_{int}={\hbar \over 2}[\Sigma_i \omega_i\sigma_z^i +\Sigma_{i<j}\pi J_{ij} \sigma_z^i \sigma_z^j]$.  To understand how the coupling will behave, we first look at the usual two body interaction.

A scalar two-body coupling propagator of the form  ${e^{-i{\pi\over 2} J_{12} \sigma_{z}^{1}\sigma_{z}^{2}t}}$ (where $\sigma_i$ are the Pauli matrices) will transform a transverse magnetization, ${\sigma_{x}^{1}}$, say, into itself and an antiphase component: 

\[
{\sigma_{x}^{1}  \stackrel{J_{12}}\longrightarrow  \sigma_{x}^{1}cos\theta + \sigma_{y}^{1}\sigma_{z}^{2}\sin\theta},
\]
where ${\theta = {\pi} J t}$.  After ${\theta = \pi /2}$, the antiphase doublet  ${\sigma_{y}^{1}\sigma_{z}^{2}}$ state is created.  An x-phase pulse on spins 1 and 2 will change this into an antiphase doublet observable on spin 2, ${-\sigma_{z}^{1}\sigma_{y}^{2}}$. 

Analogously, a three-body coupling propagator of the form  ${e^{-i{\pi\over 2} J_{123} \sigma_{z}^{1}\sigma_{z}^{2}\sigma_{z}^{3}T}}$ (where $T$ is time in the simulated system) will transform a transverse magnetization, ${\sigma_{x}^{1}}$, say, into itself and a component antiphase in the coupled spins: 

\begin{equation}
{\sigma_{x}^{2}  \stackrel{J_{123}}\longrightarrow }{\sigma_{x}^{2}cos\theta + \sigma_{z}^{1}\sigma_{y}^{2}\sigma_{z}^{3}sin\theta}. 
\label{unnatural}
\end{equation}

After ${\theta = {\pi} {J_{123}}T= \pi /2}$, the doubly antiphase quartet ${\sigma_{y}^{1}\sigma_{z}^{2}\sigma_{z}^{3}}$ is created.  
An NMR pulse sequence for implementing this evolution is straightforward to derive by geometric algebra procedures \cite{GApaper}.
The desired propagator for the three particle interaction can be expanded in terms of the available scalar couplings and free evolutions:
\begin{equation}
\ba{ll}
&e^{(-i{\pi \over 2}{J_{123}}T \sigma_z^1\sigma_z^2\sigma_z^3)}\nonumber\\

=&e^{(+i{\pi \over 4} \sigma_x^2)}
   e^{(+i{\pi \over 4}\sigma_y^2)}
   e^{(+i{\pi \over 2}{J_{123}}T \sigma_z^1\sigma_y^2\sigma_z^3)}
   e^{(-i{\pi \over 4} \sigma_y^2)}
   e^{(-i{\pi \over 4} \sigma_x^2)}\nonumber\\

=&e^{(+i{\pi \over 4} \sigma_x^2)}
   e^{(+i{\pi \over 4} \sigma_y^2)}
   e^{(+i{\pi \over 2}\sigma_z^1\sigma_x^2)} 
   e^{(+i{\pi \over 2}{J_{123}}T\sigma_z^2\sigma_z^3)} 	\times \nonumber\\ 
&  e^{(-i{\pi \over 2}\sigma_z^1\sigma_x^2)} 
   e^{(-i{\pi \over 4} \sigma_y^2)}
   e^{(-i{\pi \over 4} \sigma_x^2)}\nonumber\\
=&e^{(+i{\pi \over 4} \sigma_x^2)}
   e^{(+i{\pi \over 2} \sigma_y^2)}
   e^{(-i{\pi \over 2}\sigma_z^1\sigma_z^2)}
   e^{(-i{\pi \over 4}\sigma_y^2)} 				\times\nonumber\\  
&  e^{(+i{\pi \over 2}{J_{123}}T\sigma_z^2\sigma_z^3)}
   e^{(-i{\pi \over 4}\sigma_y^2)}  
   e^{(-i{\pi \over 2}\sigma_z^1\sigma_z^2)}
  e^{(-i{\pi \over 4} \sigma_x^2)},
\ea
\end{equation}
which implies the following pulse sequence $V_T$ to simulate the ${\sigma_{z}\sigma_{z}\sigma_{z}}$ Hamiltonian:
\begin{eqnarray}
{\left [ \pi \over 2 \right]}_{-x}^{2} \rightarrow \left [\pi \right]_{-y}^{2} \rightarrow {\left [{1\over{2J_{12}}}\right]} \rightarrow {\left [ \pi \over 2 \right]}_{y}^{2} \rightarrow \nonumber\\
{ \left [{T\over{2J_{23}}}\right]} \rightarrow {\left [ \pi \over 2 \right]}_{y}^{2} \rightarrow {\left [{1\over{2J_{12}}}\right]} \rightarrow {\left [ \pi \over 2 \right]}_{x}^{2}
\label{pulsesequence}
\end{eqnarray}
To generalize, an $m$-body interaction term can be composed of a number of two body interaction terms and single spin rotations that is linear in $m$.  If not all pairs of spins are coupled, as in the case of a linear chain, then relay gates must be used \cite{Evans} which entails only a polynomial number of additional operations.

The three quantum bit NMR system was a room temperature sample of ${\rm ^{13} C}$ labelled Alanine in deuterated water.  We identify spin 1 as the carbonyl $C$ spin, spin 2 as the $C_{\alpha}$ spin, and spin 3 as the $C_{\beta}$ spin.  The scalar couplings were $J_{12}$ = 54.2 Hz,
$J_{23}$ = 35.1 Hz, and $J_{13}$ = 1.2 Hz.  The $^{13}C$ resonance frequency at 9.4 T was 100.6 MHz, and was detected by an inverse probe.  The chemical shift difference between spins 1 and 2 was 12,580 Hz, and between spins 2 and 3 was 3443 Hz.  The proton spins were decoupled.  Initial states were prepared from the thermal equilibrium state with magnetization in all three spins by a shaped pulse that excites, for example, spins 1 and 3, followed by a magnetic field gradient that dephases the magnetization in spins 1 and 3.  Then only the spin 2 magnetization remains, which may be observed by exciting it into a transverse magnetization.  Explicitly,
\[
{\sigma_{z}^1 + \sigma_{z}^2 + \sigma_{z}^3 }  \stackrel{\left [ \pi /2 \right ]_{y}^{1,3}} \longrightarrow  {\sigma_{x}^1 + \sigma_{z}^2 + \sigma_{x}^3}  \stackrel{ grad }\longrightarrow  {\sigma_{z}^{2}}  \stackrel{\left [ \pi /2 \right ]_{y}^{2}} \longrightarrow {\sigma_{x}^{2}}.
\]
Here ${\it grad}$ refers to a magnetic field gradient, which destroys the transverse magnetization when viewed as a spatial average \cite{PhysicaD}.  

\begin{figure}
  \centering
  \epsfig{file=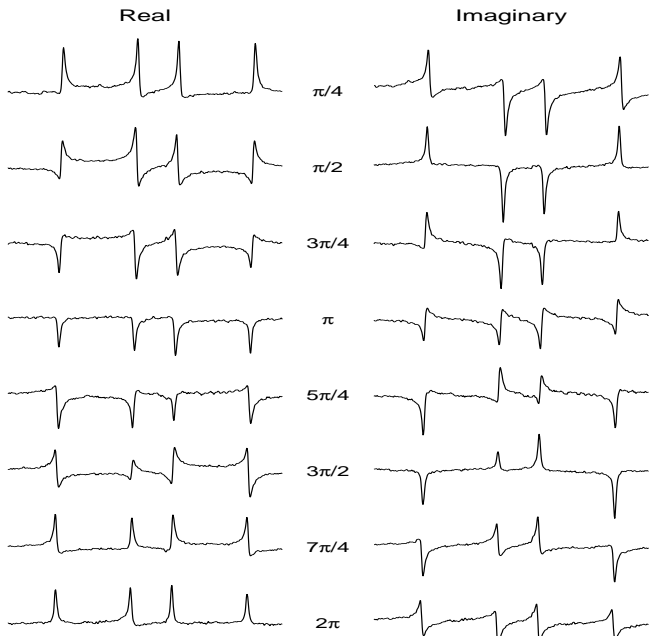, height=8.6cm, width=8.6cm}
  \caption{NMR spectra from $C_{\alpha}$ spin of $^{13}$C-labelled Alanine demonstrating a quantum 
simulation of the $\sigma_z\sigma_z\sigma_z$ Hamiltonian as implemented by the pulse sequence
\eqn{pulsesequence}. The $C_{\alpha}$ resonance is split twice by the 
couplings to the other two carbon nuclei, resulting in four lines.  As a function of the angle (or evolution time) the spectra exhibit the periodicity given by \eqn{unnatural}: $\theta = \pi /2$, doubly antiphase; $\theta = \pi$, in phase negative; $\theta = 3 \pi /2$, doubly antiphase; $\theta = 2 \pi$, back to in phase positive.  
} 
  \label{spectra}
\end{figure}

Figure \ref{spectra} shows the real and imaginary $^{13}C$ spectra observed on spin 2 for representative angles $\theta$ or delay times, confirming that the three particle propagator $e^{(-i{\pi \over 2}{J_{123}}t \sigma_z^1\sigma_z^2\sigma_z^3)}$ transforms the initial state according to Equation \ref{unnatural}.
It is clear that the spectra evolve with a periodicity $T=2/J_{123}$. The simulated time direction exhibits only one frequency (positive and negative) since there is only one possible non-zero transition energy for this system as shown in the Appendix (Eq. \ref{eigen123}).  A two-dimensional Fourier transform would then directly relate the physical eigenenergies to the simulated eigenenergies.

\section{Conclusion}
While multiple quantum coherences have been widely used in NMR, as have average Hamiltonian schemes to scale down or remove terms from the natural Hamiltonian of the physical system \cite{Warren,Levitt1,Levitt2}, the formalism described above adapts these methods to quantum information processing, in particular quantum simulation.  Using one quantum system to simulate another quantum system efficiently is a powerful idea, which may allow the study of an otherwise intractable class of problems.  Quantum simulation permits the construction of an arbitrary Hamiltonian in the simulated system,
as illustrated with the construction of the three body interaction ${\sigma_{z}\sigma_{z}\sigma_{z}}$ Hamiltonian.  The efficiency, or computational complexity, of a simulation, however, depends on how difficult it is to implement $V_T$ for a given system.  Although the results given in \cite{Lloyd2} imply that efficient approximate implementations exist for general cases of physical interest, the results discussed in this paper show that more direct, exact implementations may also be found in some cases. 

\section{Acknowledgements}
This work was supported in part by the U.S. Army Research office under
contract/grant number DAAG 55-97-1-0342 from the DARPA Ultrascale Computing
Program. R. L. thanks the National Security Agency for support.

\appendix

\section{Relation Between Eigenenergies and Product Operators}
Representations of Hamiltonians in terms of eigenenergies may be related to representations in terms of the Pauli matrices $\sigma_z^i$.  Once the arbitrary Hamiltonian is expressed in terms of many-body interactions, $\sigma_z^1\sigma_z^2\cdots\sigma_z^n$, each of these may be broken down in terms of the available external and internal Hamiltonians (two body couplings).  Multiple couplings can act at the same time, giving a possible increase in efficiency of implementation.

An arbitrary Hamiltonian, if the eigenstructure is known, can be written as
\[
{\cal H}_s \;=\; \sum_{k=0}^{\infty} {\xi}_k |\psi_k\rangle\langle \psi_k|.
\]
The $|\psi_k\rangle$ are a complete set of orthogonal states and the ${\xi}_k$ are the energy eigenvalues (not necessarily ordered by size). 
If truncated to the first $2^n$ levels we have
\[
{\cal H}_s \;=\; \sum_{k=0}^{2^n-1} {\xi}_k |\psi_k\rangle\langle \psi_k|.
\]
To simulate this with an $n$-spin system we require the map $\phi$ of Eq.\ref{scheme}. One possible map which we adopt is to map simulated eigenstates to Zeeman eigenstates:
\[
|\psi_k\rangle \;\stackrel{\phi}{\longmapsto}\; |k\rangle,
\]
where $|k\rangle$ is the binary expansion of $k$.
Other mappings are possible.  The mapping need only connect a basis set that spans the $2^n$ dimensional Hilbert spaces of the simulated and physical systems.

First we resolve the identity in terms of the idempotents $E_{\pm}^{i} = {\frac {1}{2}}(1 \pm \sigma_z^{i})$ where $\sigma_z^i$ is the Pauli spin matrix for spin $i$:
\beqnar
1 &=& (E_+^n + E_-^n) (E_+^{n-1} + E_-^{n-1})\cdots (E_+^1 + E_-^1)\nonumber\\
& = & E_+^nE_+^{n-1}\cdots E_+^1 + E_-^nE_+^{n-1}\cdots E_+^1 +\nonumber\\
&& \cdots +E_-^nE_-^{n-1}\cdots E_-^1\nonumber\\
& = & \sum_{\{\epsilon_i\}}E_{\epsilon_n}^nE_{\epsilon_{n-1}}^{n-1}\cdots E_{\epsilon_1}^1\nonumber
\eeqnar
where the $\epsilon_i$ take on the values $\pm 1$. By identifying $+1 \leftrightarrow 0$ and $-1 \leftrightarrow 1$ we may identify the {\em sequence} $\epsilon_n\epsilon_{n-1}\cdots\epsilon_2\epsilon_1$ with the binary expansion $\eta_n\eta_{n-1}\cdots\eta_2\eta_1$ of some integer $k$, where $\eta_i$ take on the values 0 or 1. A summary of the relations are:
\beqnar
\epsilon_n\epsilon_{n-1}\cdots\epsilon_2\epsilon_1 & \leftrightarrow & 
\eta_n\eta_{n-1}\cdots\eta_2\eta_1,
\eeqnar
\beqnar
\epsilon_i & = & 1-2\eta_i,\nonumber\\
k & = & \eta_12^0 + \eta_22^1 +\cdots + \eta_n2^{n-1}\nonumber\\
 & = & \sum_{i=1}^n \eta_i 2^{i-1}. \label{kexp}
\eeqnar
We may then write
\beqnar
1 & = & \sum_{\{\epsilon_i\}}E_{\epsilon_n}^nE_{\epsilon_{n-1}}^{n-1}\cdots E_{\epsilon_2}^2E_{\epsilon_1}^1 \;\; \sum_{k=0}^{2^n-1} E_{k}. 
\label{eqn3}
\eeqnar
where $ E_{k} \equiv E_{\epsilon_n^k}^nE_{\epsilon_{n-1}^k}^{n-1}\cdots E_{\epsilon_2^k}^2E_{\epsilon_1^k}^1 $. Thus we have
\beqnar
{\cal H}_s \;\stackrel{\phi}{\longmapsto}\; \bar{\cal H}_p & = & \sum_{k=0}^{2^n-1} {\xi}_k |k\rangle\langle k|\nonumber\\ 
& = & \sum_{k=0}^{2^n-1} {\xi}_k |\epsilon^k_n\epsilon^k_{n-1}\ldots\epsilon_1^k\rangle\langle\epsilon^k_n\epsilon^k_{n-1}\ldots\epsilon_1^k|\nonumber\\
 & = & \sum_{k=0}^{2^n-1} {\xi}_k E_{\epsilon^k_n}^n E_{\epsilon^k_{n-1}}^{n-1}\ldots E_{\epsilon_1^k}^1.\nonumber\\
& = & \sum_{k=0}^{2^n-1} {\xi}_k E_k.\label{dah2}
\eeqnar
In general this will have the form
\beqnar
\bar{\cal H}_p& = & \beta_0\nonumber\\
 &+& \sum_{j=1}^n\beta_j \sigma_z^n + \sum_{j<k=1}^n\beta_{jk} \sigma_z^j\sigma_z^k +\cdots+\beta_{12\ldots n}\sigma_z^1\sigma_z^2\cdots\sigma_z^n\nonumber\\
& = & \sum_{k=0}^{2^n-1}\alpha_k (\sigma_z^1)^{\eta^k_1}(\sigma_z^2)^{\eta^k_2}\cdots (\sigma_z^n)^{\eta^k_n},  \label{dah}
\eeqnar
where the $\{\alpha\}$ and $\{\beta\}$ are real numbers. Knowing the ${\xi}_k$, how do we determine the $\alpha_k$? Substitute a resolution of the identity in \eqn{dah} to get
\beqnar
\bar{\cal H}_p& = &\sum_{k=0}^{2^n-1}\alpha_k (\sigma_z^1)^{\eta^k_1}(\sigma_z^2)^{\eta^k_2}\cdots (\sigma_z^n)^{\eta^k_n} \sum_{j=0}^{2^n-1}E_j\nonumber\\
& = &\sum_{j=0}^{2^n-1}\sum_{k=0}^{2^n-1}\alpha_k (\epsilon_1^j)^{\eta^k_1}(\epsilon_2^j)^{\eta^k_2}\cdots (\epsilon_n^j)^{\eta^k_n} E_j.\label{dah3}
\eeqnar
Comparing \eqn{dah3} with \eqn{dah2} we find that
\[
{\xi}_k \;=\; \sum_{j=0}^{2^n-1} (\epsilon_1^k)^{\eta^j_1}(\epsilon_2^k)^{\eta^j_2}\cdots (\epsilon_n^k)^{\eta^j_n}\alpha_j.
\]
This may be written as a matrix equation
\beqn
\mb{\xi} \;=\; \mb{M}\mb{\alpha}
\label{efroma}
\eeqn
where
\[
\mb{M}_{kj} \;\equiv\; (\epsilon_1^k)^{\eta^j_1}(\epsilon_2^k)^{\eta^j_2}\cdots (\epsilon_n^k)^{\eta^j_n}.
\]
Noting that
\beqnar
\mb{M}\mb{M}^{\rm T} &=& \sum_{j=0}^{2^n-1} \mb{M}_{kj}\mb{M}^{\rm T}_{jl}\nonumber\\ 
& = &  \sum_{j=0}^{2^n-1} \mb{M}_{kj}\mb{M}_{lj}\nonumber\\
& = &  \sum_{j=0}^{2^n-1} (\epsilon_1^k)^{\eta^j_1}(\epsilon_2^k)^{\eta^j_2}\cdots (\epsilon_n^k)^{\eta^j_n}
(\epsilon_1^l)^{\eta^j_1}(\epsilon_2^l)^{\eta^j_2}\cdots (\epsilon_n^l)^{\eta^j_n}
\nonumber\\
& = &  \sum_{j=0}^{2^n-1} (\epsilon_1^k\epsilon_1^l)^{\eta^j_1}(\epsilon_2^k\epsilon_2^l)^{\eta^j_2}\cdots (\epsilon_n^k\epsilon_n^l)^{\eta^j_n} 
\nonumber\\
& = &  \sum_{j=0}^{2^n-1} \delta_{kl} \;=\; 2^n\delta_{kl} \;=\; 2^n \mb{1},
\eeqnar
we see that 
\[
\mb{M}^{-1} \;=\; 2^{-n}\mb{M}^{\rm T}.
\]
Thus inverting the equation \eqn{efroma} we get the $\mb{\alpha}$ in terms of the energies $\mb{\xi}$:

\beqn
\mb{\alpha} \;=\; 2^{-n}\mb{M}^{\rm T}\mb{\xi}.
\label{afrome}
\eeqn
That is
\[
{\alpha}_k \;=\; 2^{-n}\sum_{j=0}^{2^n-1} (\epsilon_1^j)^{\eta^k_1}(\epsilon_2^j)^{\eta^k_2}\cdots (\epsilon_n^j)^{\eta^k_n}{\xi}_j.
\]
The matrix $\mb{M}^{\rm T}$ is the $2^n\times 2^n$ Hadarmard matrix \cite{Hadamard}.  For example, for two spins, 
\beqnar
{\bf M}
 =
\left[
\ba{cccc}
1     &    1         &    1         &  1  \\
1     &    -1        &    1         &  -1  \\
1     &    1         &    -1        &  -1  \\
1     &    -1        &    -1        &  1  
\ea
\right].\nonumber
\eeqnar
Therefore, a simulated Hamiltonian specified by $\mb{\xi}$ determines the coefficients $\mb{\alpha}$ in the expansion of the physical Hamiltonian with terms that are of the form of many body interactions $\sigma_z^1\sigma_z^2\cdots\sigma_z^m$.  For example, the three-body interaction Hamiltonian, ${\bar{\cal H}}_p = {\hbar \over 2}\pi J_{123}\sigma_z^1\sigma_z^2\sigma_z^3$ implies $\alpha_7 = {\hbar \over 2}\pi J_{123}$, and all else is zero.  Applying Eq.~\ref{afrome} to this gives: 
\beqnar
{\mb\xi}={\hbar \over 2}\pi J_{123}{\bf diag}(+1,-1,-1,+1,-1,+1,+1,-1)
\label{eigen123}
\eeqnar

\end{document}